\documentclass{article}

\newtheorem{theorem}{Theorem}
\newtheorem{lemma}{Lemma}

\makeatletter

\@addtoreset{equation}{section}
\makeatother

 \newcommand{\fulltoday}{\number\day\space \ifcase\month\or
    January\or February\or March\or April\or May\or June\or
    July\or August\or September\or October\or November\or December\fi
    \space
\number\year}

\title{Deformed Fredkin Spin Chain with Extensive Entanglement}
\author{Olof Salberger$^{1}$, Takuma Udagawa$^{2}$, Zhao Zhang$^{3}$, \\
 Hosho Katsura$^{2}$, Israel Klich$^{3}$,  Vladimir  Korepin$^{1}$ \\
\\
\normalsize {$^{1}$C. N. Yang Institute for Theoretical Physics, Stony Brook University,} \\\normalsize {  NY 11794, USA} \\ \normalsize {$^{2}$ Department of Physics, Graduate School of Science, The University of Tokyo, } \\  \normalsize { Hongo, Tokyo 113-0033, Japan} \\ \normalsize { $^{3}$Department of Physics, University of Virginia,
Charlottesville, VA 22904, USA}
}

\date{}

\usepackage[]{graphicx}

\usepackage{amsmath}
\usepackage{braket}
\usepackage{color}

\usepackage[rm,up,sc,compact,topmarks,calcwidth,pagestyles]{titlesec}

\usepackage{latexsym}
\def\qed{\hfill $\Box$} 

\def\tm{\tilde{M}}
\def\tn{\tilde{N}}

\begin{document}
\maketitle

\begin{abstract}
We introduce a new spin chain which is a deformation of the Fredkin spin chain and has a phase transition between bounded and extensive entanglement entropy scaling. In this chain, spins have a local interaction of three nearest neighbors. 
 The Hamiltonian is frustration-free and its ground state can be described analytically as a weighted superposition of Dyck paths. 
In the purely spin $1/2$ case, the entanglement entropy obeys an area law: it is bounded from above by a constant, when the size of the block $n$ increases (and  $t>1$). When a local color degree of freedom is introduced the entanglement entropy increases linearly with the size of the block (and $t>1$).  The entanglement entropy of half of the chain is tightly bounded by ${  n}\log s$ where  $n$ is the size of the block, and  $s$ is the number of colors. 

Our chain fosters a new example for a significant boost to entropy and for the existence of the associated critical rainbow phase where the entanglement entropy scales with volume that has recently been discovered in Zhang et al. \cite{zhang2016quantum}. 
\end{abstract}

\section{Introduction}
Spin chains have been at the center of high energy physics, solid state physics, quantum optics and statistical mechanics for many years. Arguably the most famous spin chain is the Heisenberg $XXX$ chain \cite{heisenberg1928theorie}. Bethe's solution \cite{bethe1931theorie} helped to construct multiple generalizations (see e.g. \cite{korepin1997quantum}), with some generalizations playing important roles for example in the context of the AdS/CFT correspondence \cite{beisert2012review}. Some spin chains are solvable in a weaker sense: only the ground state, and perhaps a few other eigenstates, can be described analytically. The best-known example for such a chain is the AKLT chain \cite{affleck1987rigorous,affleck1988valence}.  Chains where the ground state can be efficiently described are often ``frustration free'' in that the ground state is a common ground state of the individual local terms comprising the Hamiltonian. 

\par
We are searching for solvable spin chains with a high level of quantum fluctuations.  Such fluctuations manifest themselves in the entropy of subsystems even when the complete state of the system is known, in contrast with classical systems. 
  Indeed, in a classical theory of random variables, a zero entropy state implies that the probability measure is concentrated on one particular realization. Thus all local variables have definite values, and in particular, there is no entropy in any sub-system.  This is not true in the quantum case, where a sub-system can have enhanced entropy if strong quantum fluctuations are present even under the condition that the total entropy is zero. 

For spin chains, the usual setup is to find a Hamiltonian with a unique ground state, ensuring that the entropy of the total quantum state is zero. Let us consider a block of spins [in the ground state] of length $n$ in a spin-chain with only local interactions. How does the entropy of the block behave as $n$ increases? For gap-full spin-chains in 1D the entropy is bounded \cite{hastings2007area} as $n \rightarrow \infty$, this behavior is conjectured to be typical also in higher dimensional systems, and is often referred to as an area law scaling (for a review see, e.g. \cite{eisert2010colloquium}). For many gapless chains the entropy increases logarithmically with the growth of $n$, which is the typical behavior for conformal field theories \cite{holzhey1994high,calabrese2009entanglement}. Logarithmic violations of area law behavior are also typical for gapless fermion models in higher dimensions \cite{gioev2006entanglement,wolf2006violation}. 

 However, it has been recently realized that entropy can grow much faster: indeed as fast as the (maximally possible) volume scaling. One way to achieve this is, of course, allowing for non-local terms in the Hamiltonian that can generate long-range correlations, as studied, e.g. in \cite{gori2015explicit}). However, most physical systems are well described by local Hamiltonians, thus we stress the fact that we are considering here strictly local spin-chains with enhanced entropy. Even within the restriction of strict locality, a special spin chain with entropy scaling as fast as volume was found in \cite{irani2010ground}. In addition, particular examples of systems with volume scaling, but with Hamiltonians that are not translationally invariant have been presented in \cite{gottesman2010entanglement,vitagliano2010volume,ramirez2014conformal}.

Recently, the existence of an entire quantum critical phase, featuring extensive entanglement has been demonstrated for a class of spin-chains \cite{zhang2016quantum}. The phase is characterized by a linear scaling of entropy and has a phase transition into an area law regime. With respect to certain variables (color), the ground state may be thought of as a "rainbow" state where pairs of spins diametrically opposite the middle of the chain are strongly entangled. Such a state has been discussed in terms of tensor network states and even produced in the laboratory using photonics states entangled on the frequency comb \cite{chen2014experimental}, moreover, it has been shown that this type of state may be generated in a free fermion model with a local Hamiltonian \cite{vitagliano2010volume,ramirez2014conformal}. The free fermion model is very useful as direct numerical simulations are possible. However, unfortunately, the coupling constants are not translationally invariant: in particular, this means that when going to a large system limit, new sites are added to the system with couplings that must be explicitly tuned, and are different from the couplings already existing in the bulk of the original system. This limitation makes it hard to discuss quantum phases, as even at finite temperatures thermodynamic quantities such as free energy would not scale with the volume as they would for any translationally invariant system.

The transition point itself (area law and extensive entropy), is very interesting and described by the Motzkin spin chain model of \cite{Movassagh07112016}. It is novel as it is different from many of the quantum critical points studied before (which are most often described by conformal field theories) and has entropy scaling as the square root of the volume. In this model the ground state is represented by a superposition of so-called colored Motzkin paths and the extensive entropy phase of \cite{zhang2016quantum} is obtained by deforming the Hamiltonian to have a particular weighted type of superposition favoring highly entangled states.

Another square root enhancement of entanglement has also been recently demonstrated in the Fredkin spin chain of \cite{salberger2016fredkin,dell2016violation}. The Fredkin spin chain is a direct generalization of the $XXX$ spin chain: neighboring spins switch on and off the Heisenberg interaction (see formula (3) of  \cite{salberger2016fredkin}).
The Fredkin chain has a unique ground state: it is a uniform superposition of Dyck paths. Much as in the Motzkin chain of \cite{Movassagh07112016,bravyi2012criticality}, the colorless case features entanglement entropy scales as $\log n$ while in the colored case the entropy scales as $\sqrt n$. 

Here we introduce a deformed Fredkin chain that is a new example of the extensively entangled critical phase of \cite{zhang2016quantum}, showing that this is a feature in a larger class of systems. Owing to the 3-term Fredkin interaction, the treatment of the ground state involves Dyck paths rather than Motzkin paths (appearing in \cite{zhang2016quantum}) and makes the analysis simpler to carry out. The coefficient of the volume law scaling is enhanced compared to the deformed Motzkin chain of \cite{zhang2016quantum} from $\log({d-1\over 2})$ to $\log({d\over 2})$ where $d$ is the local Hilbert space dimension. In particular, the counting of Dyck paths that we use is intimately related to interesting problems whose combinatorics have been extensively studied, namely the enumeration of Young diagrams.

\section{Hamiltonian and ground state}
We first explain the relationship between Dyck path and spin $1/2$ chains. A Dyck path on $2n$ steps is any path from $(0, 0)$ to $(0, 2n)$ with steps $(1, 1)$ and $(1, -1)$ that never passes below the $x$-axis. When we regard a $(1, 1)$ step a local up-spin as and a $(1, -1)$ step as a local down-spin, the Dyck path corresponds to a state of $2n$ spins. The unique ground state of a Fredkin spin chain is a uniform superposition of Dyck paths. When the $j$th step is assigned a color $c_j$, picked among a set of $s$ colors, $\left|\uparrow^{c_j}\right\rangle$ is  a spin $+c_j/2$ state. In this way, the local Hilbert space dimension is $2s$. The ground state of the Fredkin model with a color degree of freedom is a uniform superposition of colored Dyck paths \cite{salberger2016fredkin,dell2016violation}.

We introduce the Hamiltonian of a deformed Fredkin spin chain with the parameter $t$ while remaining frustration free. The Hamiltonian is
\begin{eqnarray}
H(s,t)=H_{F}(s,t)+H_{X}(s)+H_{\partial}(s).  \label{Hamiltonian}
\end{eqnarray}
Here 
\begin{eqnarray}
H_{F}(s,t)=\sum_{j=2}^{2n-1}\sum_{c_1,c_2,c_3=1}^{s}\big(|\phi^{{c_{1},c_{2},c_{3}}}_{j,+}\rangle\langle\phi^{{c_{1},c_{2},c_{3}}}_{j,+}|+|\phi^{{c_{1},c_{2},c_{3}}}_{j,-}\rangle\langle\phi^{{c_{1},c_{2},c_{3}}}_{j,-}|\big) ~,\label{HF}
\end{eqnarray}
with
\begin{eqnarray}
|\phi^{{c_{1},c_{2},c_{3}}}_{j,+}\rangle=\frac{1}{\sqrt{1+t^2}}(\left|\uparrow_{j-1}^{c_1}\uparrow_j^{c_2}\downarrow_{j+1}^{c_3}\right\rangle-t\left|\uparrow_{j-1}^{c_2}\downarrow_j^{c_3}\uparrow_{j+1}^{c_1}\right\rangle)~,
\end{eqnarray}
\begin{eqnarray}
|\phi^{{c_{1},c_{2},c_{3}}}_{j,-}\rangle=\frac{1}{\sqrt{1+t^2}}(\left|\uparrow_{j-1}^{c_1}\downarrow_j^{c_2}\downarrow_{j+1}^{c_3}\right\rangle-t\left|\downarrow_{j-1}^{c_3}\uparrow_j^{c_1}\downarrow_{j+1}^{c_2}\right\rangle)~,\label{phiminus}
\end{eqnarray}
and  
\begin{eqnarray}
H_X(s)=&\sum_{j=1}^{2n-1}[\sum_{c_1\neq c_2}\left|\uparrow_j^{c_1}\downarrow_{j+1}^{c_2}\right\rangle\left\langle\uparrow_j^{c_1}\downarrow_{j+1}^{c_2}\right| \nonumber\\
+&\frac{1}{2}\sum_{c_1,c_2=1}^{s}(\left|\uparrow_j^{c_1}\downarrow_{j+1}^{c_1}\right\rangle-\left|\uparrow_j^{c_2}\downarrow_{j+1}^{c_2}\right\rangle)(\left\langle\uparrow_j^{c_1}\downarrow_{j+1}^{c_1}\right|-\left\langle\uparrow_j^{c_2}\downarrow_{j+1}^{c_2}\right|)].
\end{eqnarray}
  The boundary conditions, that the first step in a Dyck path is upwards and the last step is  downwards, are implemented by:  
\begin{eqnarray}
H_{\partial}(s)=\sum_{c=1}^{s}\left|\downarrow_1^c\right\rangle\left\langle\downarrow_{1}^c\right|+\left|\uparrow_{2n}^c\right\rangle\left\langle\uparrow_{2n}^c\right|.
\end{eqnarray}
  Note that we are using here the parameter $t$ as a deformation parameter, as opposed to $q$ which is often used to denote deformations. In what comes next we always take $t$ real. 

The unique ground state of this model is
\begin{eqnarray}
\left|GS\right\rangle=\frac{1}{\mathcal{N}}\sum_{w\in\{s-colored\,Dyck\,paths\}}t^{{1\over 2}\mathcal{A}(w)}\left|w\right\rangle~,\label{GS} 
\end{eqnarray}
where $\mathcal{N}$ is a normalization factor and $\mathcal{A}$ is the area below the path $w$. 
  That \eqref{GS} is the ground state can be verified directly, namely, by checking that $\left|GS\right\rangle$ is annihilated by each of the projection operators making up the deformed Fredkin Hamiltonian.

\begin{figure} \centering \includegraphics[width=0.4\textwidth]{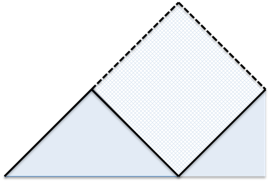} \caption{  The difference in the area under an up-up-down Dyck path and an up-down-up Dyck path is 2 (in units where the spin chain length is 2n).  } \label{AreaChange} \end{figure}

For example, the area under a Dyck path containing the spin configuration $|....\uparrow_{j-1}^{c_1}\uparrow_j^{c_2}\downarrow_{j+1}^{c_3}....\rangle$, will be larger than the area under the same Dyck path where the $j,j+1$ spins have been interchanged   (getting
$|....\uparrow_{j-1}^{c_2}\downarrow_j^{c_3}\uparrow_{j+1}^{c_1}...\rangle$) by exactly $2$, as illustrated in Fig. \ref{AreaChange}. Thus, in the ground state superposition, \eqref{GS} the relative amplitude between these will be $$t^{\text{area difference/2}}=t.$$ The states will thus be annihilated by the projectors $|\phi^{{c_{1},c_{2},c_{3}}}_{j,+}\rangle$  in $H_{F}$, \eqref{HF}.
In general, every time a pair of $\uparrow\downarrow$ is moved around a third neighbor, the area below the Dyck walk changes by a unit of $2$, while the weight of it changes by a factor of $t$.  Meanwhile, the projections in $H_{X}$, are automatically satisfied whenever the colorings obey the rules of colored Dyck paths.

Therefore the zero energy ground state (GS) of the Hamiltonian is indeed a superposition of the Dyck walks weighted by $t$ to the power of the area $\mathcal{A}$ below them (which also satisfy the boundary projectors). 
It is also straightforward to see that the ground state is unique: any other superposition of walks will violate at least one of the projectors and will have non-zero energy.  

To study entanglement of half a chain with the rest, we rewrite the ground state in the Schmidt decomposition form
\begin{eqnarray}
\left|GS\right\rangle=\sum_{m=0}^{n}\sqrt{p_{n,m}(s,t)}\sum_{x\in\{\uparrow^1,\uparrow^2,\cdots,\uparrow^s\}^m}\left|\hat{C}_{0,m,x}\right\rangle_{1,\cdots,n}\otimes\left|\hat{C}_{m,0,\bar{x}}\right\rangle_{n+1,\cdots,2n},
\end{eqnarray}
where
\begin{eqnarray}
&p_{n,m}(s,t)=\frac{M_{n,m}^2(s,t)}{N_n(s,t)},\label{pnm}\\
&M_{n.m}(s,t)=s^{\frac{n-m}{2}}\sum_{w\in\{1st\, half\,of\,Dyck\,paths\atop stopping\,at\,(n,m)\}}t^{\mathcal{A}(w)},\label{defMnm}\\
&N_n(s,t)=\sum_{m=0}^{n}s^m M_{n,m}^2(s,t)=M_{2n,0}(s,t)
\end{eqnarray}
and $\left|\hat{C}_{a,b,x}\right\rangle$ is a weighted superposition of spin configuration with $a$ excess $\downarrow$, $b$ excess $\uparrow$ and a particular coloring $x$ of unmatched arrows.

With these definitions the entanglement entropy of half a chain is given by:
\begin{eqnarray}
S_n=-\sum_{m=0}^{n}s^{m}p_{n,m}(s,t)\log p_{n,m}(s,t).
\end{eqnarray}

\section{Entanglement entropy}

\subsection{Colorless model: $s=1,\,t>1$}
  We start with a deformation of the colorless Fredkin model, namely the case of $t>1$ and $s=1$. We define $N\equiv n-m$.
\begin{lemma}\label{lem1}
When $t>1$ and $s=1$, $p_{n,m}(s,t) (=p_{n,n-N})$ satisfies the following inequality and equation
\begin{eqnarray}
\frac{t^{-\frac{1}{2}N^2}}{C(t)}< p_{n,n-N}< t^{-\frac{1}{2}N^2}C(t)^2\,\,\,\,\,\text{ for even }\,N,
\end{eqnarray}
\begin{eqnarray}
p_{n,n-N}=0\,\,\,\,\,\text{ for odd }\,N\,\label{oddN}
\end{eqnarray}
where $C(t)$ is an $n$ independent constant.
\end{lemma}

\begin{figure}
\begin{center}
\includegraphics[width=8cm,clip]{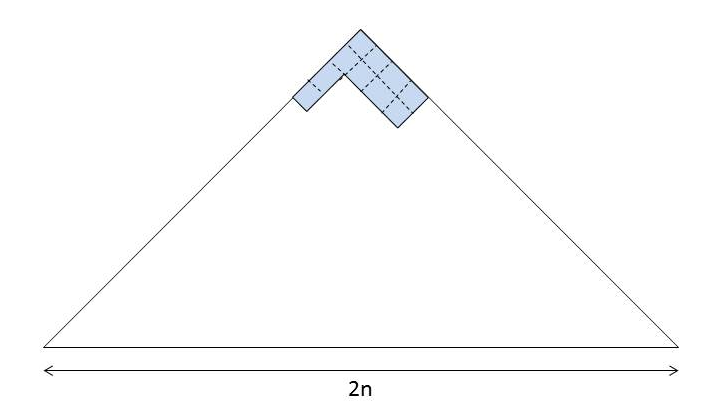}
\caption{The blue area can be regarded as a Young diagram, so the coefficient of $t^{n^2-2k}$ in Eq.\,(\ref{1}) is the partition function $P(k)$ when $k\leq n-1$. However, when $k\geq n$, the coefficient of $t^{n^2-2k}$ is less than $P(k)$ because the area where we can make the Young diagram is restricted.} \label{fig1}
\end{center}
\end{figure}

\begin{figure}
\begin{center}
\includegraphics[width=7cm,clip]{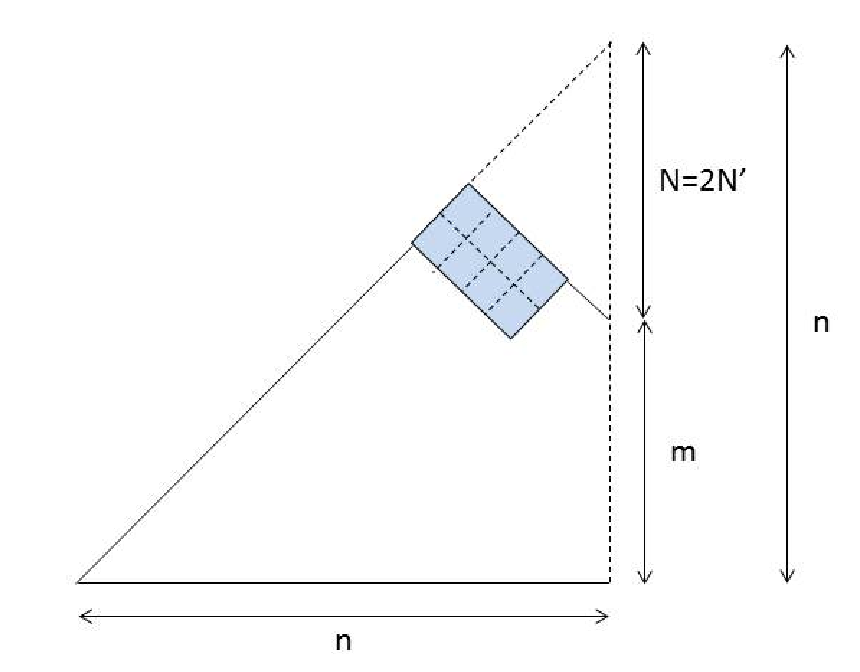}
\caption{A Dyck path reaching $(n,n-N)$ and the associated Young diagram.} \label{fig2}
\end{center}
\end{figure}

$Proof.$\\
We first note that there is no path which stops at $(n,n-N)$ where $N$ is odd. Therefore, $p_{n,n-N}=0$ for odd $N$, Eq.\eqref{oddN}. 

Next, we consider the case of even $N$.   Writing explicitly the normalization factor  
\begin{eqnarray}
N_n(s=1,t)=M_{2n,0}(s=1,t)=t^{n^2}+t^{n^2-2}+2t^{n^2-4}+3t^{n^2-6}+\cdots+t^n. \label{1} 
\end{eqnarray}
  The coefficient of a term like $t^{n^{2}-2k}$ in Eq.\,(\ref{1}) is the number of Dyck paths of area $n^{2}-2k$.  We note that:  
\begin{eqnarray}
N_n(s=1,t)&=&t^{n^2}+t^{n^2-2}+2t^{n^2-4}+3t^{n^2-6}+\cdots+t^n\\
&<& t^{n^2}\sum_{k=0}^{\infty}t^{-2k}P(k)\equiv t^{n^{2}}C(t),
\end{eqnarray}
where $P(k)$ is the integer partition of $k$ as illustrated in Fig. \ref{fig1}. More precisely, $P(k)$ is the number of Young diagrams made up of $k$ boxes. $C(t)=\sum_{k=0}^{\infty}t^{-2k}P(k)$ is a generating function of integer partitions and is known to converge when $t>1$ 
\cite{andrews2004integer}. 

Note that $N_n(t)>t^{n^2}$, and we thus have
\begin{eqnarray}
t^{n^2}<N_n(s=1,t>1)<t^{n^2}C(t). \label{2}
\end{eqnarray}
  Dyck paths reaching a height $(n,n-N)$, can have at most area ${1\over 2}n^{2}-{1\over 4}N^{2}$ (Obtained by going up $n-{N\over 2}$ steps and then making  ${N\over 2}$ downward steps). As Fig.\ref{fig2} shows, these paths correspond to partitions of $n^{2}-{1\over 2}N^{2}-2k$ where $k$ is the number of greyed boxes in Fig. \ref{fig2}. Therefore we have:  
\begin{eqnarray}
M_{n,m}(s=1,t>1)^2&=&M_{n,n-N}(s=1,t>1)^2=(\,t^{\frac{1}{2}n^2-\frac{1}{4}N^2}
+ t^{\frac{1}{2}n^2-\frac{1}{4}N^2-2} \nonumber \\
&+&2t^{\frac{1}{2}n^2-\frac{1}{4}N^2-4}+\cdots t^{\frac{1}{2}n^2-nN+\frac{1}{2}N^2+\frac{1}{2}N}\,)^2 \\ &<& t^{-\frac{1}{2}N^2}t^{n^2}\big(\, \sum_{k=0}^{\infty}t^{-2k}P(k)\,\big)^2
=t^{-\frac{1}{2}N^2}t^{n^2}C(t)^2.
\end{eqnarray}
Therefore,
\begin{eqnarray}
t^{-\frac{1}{2}N^2}t^{n^2}\leq M_{n,m}(s=1,t>1)^2<t^{-\frac{1}{2}N^2}t^{n^2}C(t)^2. \label{3}
\end{eqnarray}
Combining Eq.\,(\ref{2}) and Eq.\,(\ref{3}), with the definition \eqref{pnm} we get the following inequality
\begin{eqnarray}
\frac{t^{-\frac{1}{2}N^2}}{C(t)}< p_{n,n-N}< t^{-\frac{1}{2}N^2}C(t)^2.
\end{eqnarray}
\qed

This lemma will be used to prove that the entanglement entropy of half a chain is bounded:
\begin{theorem}
When $t>1$ and $s=1$, there exists an $n$ independent constant $D_1(t)$ such that the entanglement entropy $S_n$ satisfies $S_n(s=1,t>1)<D_1(t)$.
\end{theorem}

$Proof.$
\begin{eqnarray}
S_n(s=1,t>1)&=&-\sum_{m=0}^{n}p_{n,m}\log p_{n,m}\\
&=&-\sum_{N=0}^{n}p_{n,n-N}\log p_{n,n-N}\\
&=&-\sum_{N'=0}^{\lfloor \frac{n}{2} \rfloor}p_{n,n-2N'}\log p_{n,n-2N'}\label{554} \\
&<&-\sum_{N'=0}^{\lfloor \frac{n}{2} \rfloor}t^{-2N'^2}C(t)^2\log p_{n,n-2N'} \label{111}\\
&<&-\sum_{N'=0}^{\lfloor \frac{n}{2} \rfloor}t^{-2N'^2}C(t)^2\log (t^{-2N'^2}\frac{1}{C(t)}) \label{999}\\
&<&-\sum_{N'=0}^{\infty}t^{-2N'^2}C(t)^2\log (t^{-2N'^2}\frac{1}{C(t)}) \\
&=&C(t)^2\sum_{N'=0}^{\infty}(2N'^2\log t +\log C(t))t^{-2N'^2} \label{09}\\
&\equiv&D_1(t)
\end{eqnarray}
We defined $N=2N'$ and used lemma 1 in Eq.\,(\ref{554}), Eq.\,(\ref{111}) and Eq.\,(\ref{999}). Since the first and the second terms in Eq.\,(\ref{09}) are convergent, $D_1(t)$ is an $n$ independent constant. 
\qed

$Remark$. $D_1(t)$ is roughly estimated by $$D_1(t)\sim C(t)^2\!\!\int_0^{\infty}dN' [2N'^2\log t +\log C(t)]t^{-2N'^2}=\frac{C(t)^2}{2}\sqrt{\frac{\pi}{2\log t}}(\log C(t) +\frac{1}{2}).$$
  Since $C(t)\rightarrow 1$ when $t\rightarrow \infty$, we have $D_{1}\rightarrow 0$, and $S_{n}\rightarrow 0$. In this limit only the contribution from the dominant path, with area $n^{2}$ is important. Since there is only one ``highest area'' path, corresponding to the spin configuration $|\uparrow_{1}....\uparrow_{n}\downarrow_{n+1}...\downarrow_{2n}\rangle$, the system is close to being in a product state and clearly not entangled.  

\subsection{Colorful model: $s>1,\,t>1$}
  We now add the color degree of freedom, and use   $N$, $N'$ and $C(t)$ as defined in the previous subsection.   
It is useful to relate the probabilities of the colored case to those of the non-colored. To do so, note that using the definition, Eq. \eqref{defMnm}, we have :
\begin{eqnarray}
M_{n.m}(s,t)=s^{\frac{n-m}{2}}\sum_{w\in\{1st\, half\,of\,Dyck\,paths\atop stopping\,at\,(n,m)\}}t^{\mathcal{A}(w)}
=s^{N'}M_{n,m}(s=1,t),
\end{eqnarray}
and 
\begin{eqnarray}
N_n(s,t)=M_{2n,0}(s,t)=s^n N_n(s=1,t).
\end{eqnarray}
In particular, the colored probabilities $p_{n,m}(s,t)$ are related to the uncolored ones by:
\begin{eqnarray}
p_{n,m}(s,t)=\frac{M_{n,m}^2(s,t)}{N_n(s,t)}=s^{-m}p_{n,m}(s=1,t).\label{098}
\end{eqnarray}
We are now ready to address the entropy in the colored case. 
Eq. \eqref{098} allows us to write $S_{n}(s,t)$ explicitly in terms of the probabilities $p_{n,m}(1,t)$ of the uncolored case, yielding:
\begin{eqnarray}
S_n(s,t)&=&-\sum_{m=0}^{n}s^{m}p_{n,m}(s,t)\log p_{n,m}(s,t)\\ &=& -\sum_{m=0}^{n}p_{n,m}(1,t)(\log p_{n,m}(1,t)-m \log s)\label{entRel00}\\ &=& S_{n}(1,t)+\log s \sum_{m=0}^{n}m p_{n,m}(1,t). \label{entropyGen}
\end{eqnarray}
For $t>1$, we have established in Theorem 1 that $S_{n}(1,t)$ is bounded. Thus the behavior of the entropy is determined by the remainder term in the last equation. We find that:

\begin{theorem}
When $t>1$ and $s>1$, the entanglement entropy of ground state $S_n$ satisfies the inequality
\begin{eqnarray}
n\log s+S_n(1,t) -D_2(s,t) < S_n(s,t) < n\log s + S_n(1,t)\label{thm02}
\end{eqnarray}
where $D_2(s,t)$ is an $n$ independent constant.
\end{theorem}
As we have seen before, $S_n(1,t)<D_{1}(t)$, the above theorem shows that $S_n(s,t)=n \log s+O(1)$.

$Proof.$\\
As before, we recalling that $n-m$ is even and writing $m=n-2N'$, Eq. \eqref{entropyGen} gives us:
\begin{eqnarray}
S_n(s,t)&=& S_{n}(1,t)+\log s \sum_{m=0}^{n}m p_{n,m}(1,t)\\ 
&=& S_{n}(1,t)+\log s \sum_{N'=0}^{\lfloor \frac{n}{2} \rfloor} p_{n,n-2N'}(1,t)(n-2N')
\\&=& S_{n}(1,t)+n\log s -2\log s \sum_{N'=0}^{\lfloor \frac{n}{2} \rfloor} p_{n,n-2N'}(1,t)N'.\label{lasteq}
\end{eqnarray}
Using Lemma \ref{lem1} we can immediately bound the remainder sum in \eqref{lasteq}:
\begin{eqnarray*}
0<2\log s \sum_{N'=0}^{\lfloor \frac{n}{2} \rfloor} p_{n,n-2N'}(1,t)N'
&<&2C(t)^2\log s \sum_{N'=0}^{\lfloor \frac{n}{2} \rfloor} t^{-\frac{1}{2}N^2} N' \\
&<&2C(t)^2\log s \sum_{N'=0}^{\infty}t^{-2N'^2}N'\equiv D_2,
\end{eqnarray*}
where $D_2$ is an $n$ independent constant since the sum $\sum_{N'=0}^{\infty}t^{-2N'^2}N'$ is convergent for $t>1$. This proves the extensivity result for the entropy Eq. \eqref{thm02}.
\qed

$Remark$. $D_2(s,t)$ is roughly estimated by $$D_2(s,t)\sim\int_0^{\infty}dN'\,2C(t)^2t^{-2N'^2}N'\log s =C(t)^2\log s/2\log t.$$

\subsection{Both Colorless and colorful models, $s \geq 1,\,t<1$.}

\begin{figure}
  \centering
 \includegraphics[scale = 0.65]{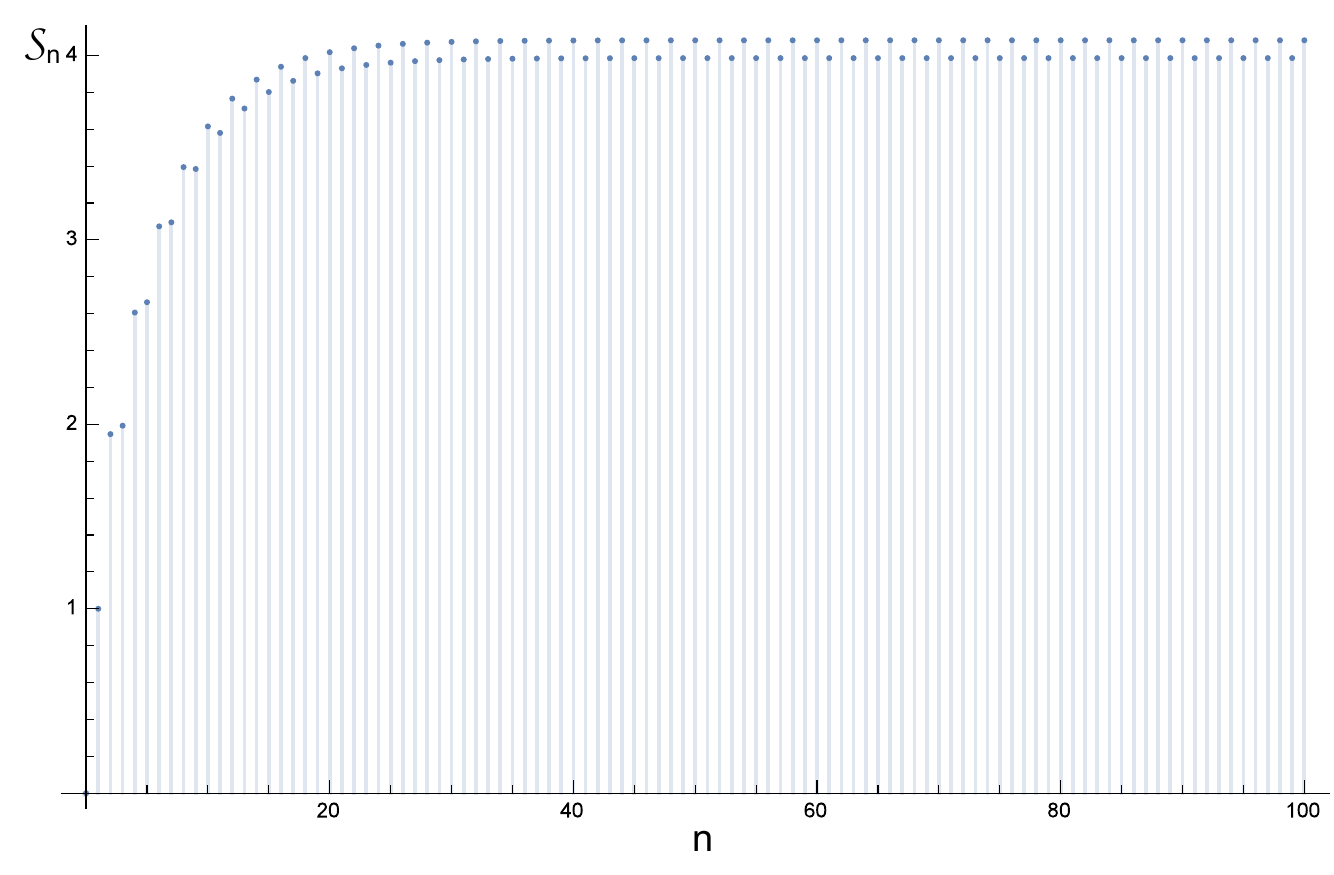}
 \includegraphics[scale = 0.65]{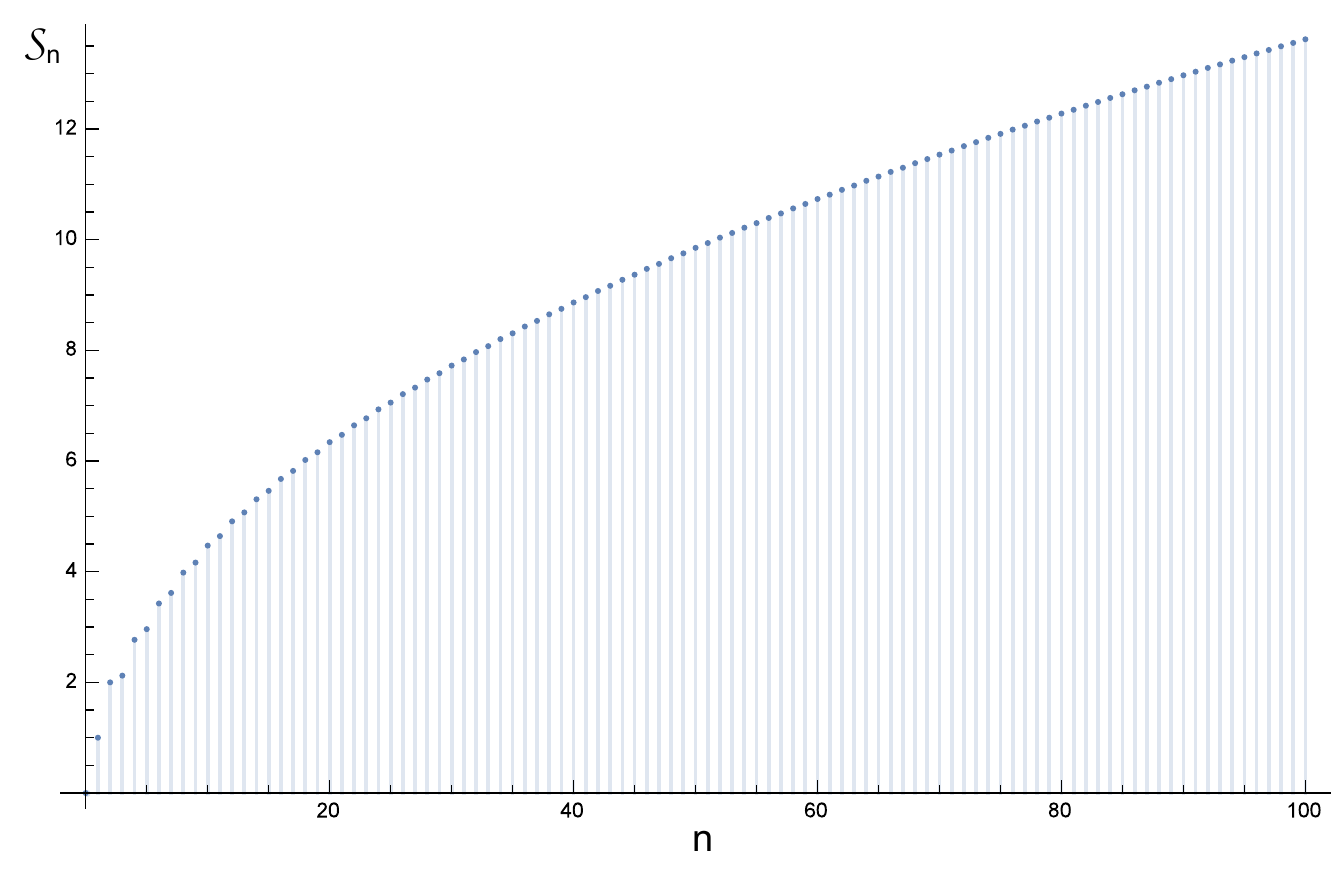}
 \includegraphics[scale = 0.65]{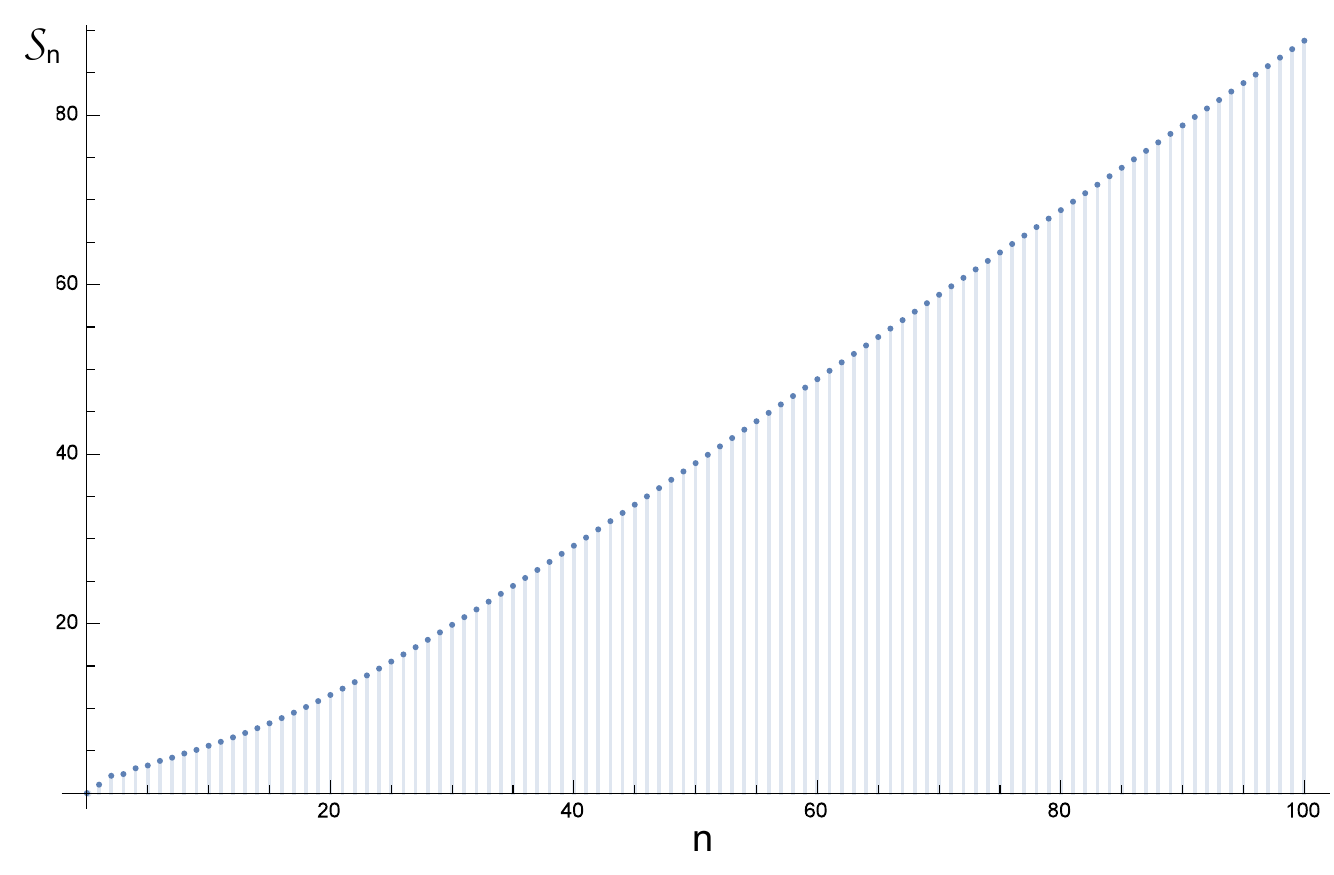}
 \caption[Entropies]{Here, we present the dramatic change in entropy as we go through the phase transition point $t=1$. Plots of exact values of the half-chain entanglement entropy as function of half chain length $n$, with $s = 2$ and with $t = {0.95 , 1.00 , 1.05}$ are shown. $S_n<4$ when $t=0.95$, $S_n\propto c\sqrt{n}$ ($c\sim 1.3$) when $t=$ and $S_n\propto n\log 2$ for $t=1.05$. }
  \label{fig:entropies}
\end{figure}

In this case, the model has bounded entropy.  The area law can be viewed as a consequence of the fact that $t<1$ exponentially favors paths with lowest possible area. The superposition of lowest height paths, can be written in the form $$|(\sum_{c_1=1}^s\uparrow_1^{c_1}\downarrow_2^{c_1})(\sum_{c_2=1}^s\uparrow_3^{c_2}\downarrow_4^{c_2})..(\sum_{c_n=1}^s\uparrow_{2n-1}^{c_n}\downarrow_{2n}^{c_n})\rangle.$$
Therefore, these paths carry only nearest-neighour non trivial correlations. Longer range entanglement must come from higher paths. While such paths are more numerous, their amplitude is  exponentially suppressed: paths that produce correlations on a distance $d$, say between the spins at site $j$ and $j+d$ must stay above the height where they first encountered $j$ until they reach $j+d$ such paths have an area at least $2d-1$ larger than the lowest area paths, and appear with a relative amplitude of at least $t^d$ in the ground states. While this argument is intuitively appealing, still, one has to properly account for the fact that such paths are numerous. Below, we establish the boundedness of the entropy for $t<1$.  

\begin{figure}
  \centering
 \includegraphics[scale = 1.6]{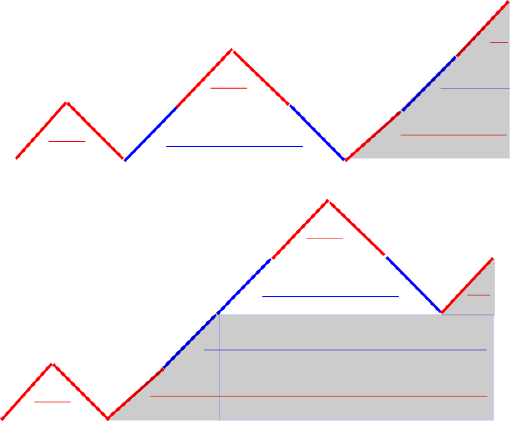}
  \caption[Change of area below path as unmatched steps are moved.]{The central idea behind the $t<1$ case. Moving unmatched steps away from their rightmost position at the edge forces the area below the path to increase. When paths are weighted by $t^A$, the weighted sum of paths can be bounded above by a geometric sum. The convergence of the geometric sum causes an $n$-independent and thus constant upper bound to exist.}
  \label{fig:geomsum}
\end{figure}

For convenience, we make slight changes in notation and introduce the symbols $\tilde{M}$, $\tilde{N}$, defined as:
\begin{eqnarray}
\tm_{n,m}(t) = t^{-\frac{n}{2}} M_{n,m}(1,t), \\
\tn_n(t) = \tm_{2n,0},
\end{eqnarray}
which corresponds to a change in normalization of $\ket{\hat{C}_{a,b,x}}$ so that the coefficient of the basis state corresponding to the lowest area path is 1. The analog of \eqref{1} is then:

\begin{eqnarray}
\tn_n(t)=\tm_{2n,0}(t)=t^{n^2-n}+t^{n^2-2-n}+2t^{n^2-4-n}+3t^{n^2-6-n}+\cdots+1. \label{renorm} 
\end{eqnarray}

This change in normalization ensures that $\tn_n$ is a strictly increasing function of $n$ for arbitrarily small $t$. This can be shown by observing that $\tn_n$ obeys the Catalan-like recursion relation $\tn_{n+1} = \sum_{k=0}^n t^{2k} \tn_k \tn_{n-k} $, with $\tn_0 = 1$. 

\begin{lemma}
When $t<1$, we have that $\tm_{n,m}(t)$ is bounded by
\begin{eqnarray}
  \tm_{n,m}(t<1) \leq \tm_{n-m,0} \frac{t^{\frac{m(m-1)}{2}}}{(1-t)^m}~~.
\end{eqnarray}
\end{lemma}
$Proof.$\\
The key idea here is that the paths counted by $\tm_{n,m}(s,t)$ can be
constructed by inserting unmatched steps into a Dyck path of length $n-m$.
Inserting an unmatched step at a distance d from the edge increases the area
by $\frac{d}{2} + \frac{1}{4}$. This leads to a term with weight $t^d$ in $\tm$, where we note that the definition of $\tm$ earlier in this section conveniently eliminates $t^{\frac{1}{2}}$ factors. Thus we can bound $\tm_{n,m}(s,t)$ from above with
\begin{eqnarray}
  \tm_{n,m}(s,t) \leq \tm_{n-m,0}(s,t) \prod_{k=0}^{m-1} \sum_{j = k}^{\infty} t^{d} = \\
  \tm_{n-m,0}(s,t) \prod_{k=0}^{m-1} \frac{t^k}{1-t} =   \tm_{n-m,0} \frac{t^{\frac{m(m-1)}{2}}}{(1-t)^m}
\end{eqnarray} 
\qed

\begin{lemma}
  For any $t < 1$, we have that
  \begin{eqnarray}
    \label{eq:finebound}
    p_{n,m}(1,t) \leq \frac{t^{m(m-1)}}{(1-t)^{2m}}
  \end{eqnarray}
\end{lemma}
$Proof.$\\

We use the definition of $p_{n,m}$ ,namely
\begin{eqnarray}
  p_{n,m} = \frac{M_{n,m}^2}{N_{n}} = \frac{M_{n,m}^2}{M_{2n,0}} = \frac{\tm_{n,m}^2}{\tm_{2n,0}}
\end{eqnarray}
 where in the last step we observe that the factors of $t$ cancel out. We may then insert the approximation in the previous lemma, which gives us
\begin{eqnarray}
  p_{n,m} = \frac{\tm_{n,m}^2}{\tm_{2n,0}} \leq \frac{\tm_{n-m,0}^2}{\tm_{2m,0}} \frac{t^{(m-1)m}}{(1-t)^{2m}} 
\end{eqnarray}
Our next step is then simply to observe that since $\tm_{2k,0} = \tn_{k} $  is a monotonously increasing function of $k$ and $p_{n,m}$ is nonzero only for even $n-m$, $ \frac{\tm_{n-m,0}^2}{\tm_{2m,0}} \leq
\frac{\tm_{n,0}^2}{\tm_{2m,0}}$. However, the RHS in this last inequality is simply
$p_{n,0} \leq 1$, meaning that we can write down our final upper bound:
\begin{eqnarray}
    p_{n,m}(1,t) \leq \frac{t^{m(m-1)}}{(1-t)^{2m}},
\end{eqnarray}
which is fully independent of $n$. \qed 

\begin{lemma}\label{lem04}
  For    $0<t < 1$, there exists an $m_0(t)$ such that
  \begin{equation}
    \label{eq:coarsebound}
    p_{n,m}(1,t) \leq
    \begin{cases}
      1 & m < m_0(t)  \\
      e^{-m} & m \geq m_0(t)
    \end{cases}
  \end{equation}
\end{lemma}

$Proof.$\\
To show this, we use the fact that all $p_{n,m} \leq 1$, combined with a coarse upper bound for the approximation in
\eqref{eq:finebound}. That is, we observe that for any $t < 1 $ , there exists a
$m_0$ such that for all $m > m_0$,
\begin{eqnarray}
  \frac{t^{m-1}}{(1-t)^2} < \frac{1}{e}.
\end{eqnarray}
The bound \eqref{eq:finebound} then implies that for $m>m_{0}$ we have:
\begin{eqnarray}
    p_{n,m}(1,t) \leq \left(\frac{t^{(m-1)}}{(1-t)^{2}}\right)^{m} < e^{-m}~~.
\end{eqnarray}
\qed

\begin{theorem}
  For any $t < 1 $ and s, there exists a constant $C(t,s)$ such that the entropy of
  the ground state $S_n$ satisfies $S_n < C$ for all n.
\end{theorem}

$Proof.$\\
To show this, we make use of the Lemma \ref{lem04} above and Eq. \eqref{entRel00}. We write  $S_n$ as
\begin{eqnarray*}
  S_n &=& \sum_{m=0}^{n} [ p_{n,m}(1,t) m \log s - p_{n,m}(1,t) \log(p_{n,m}(1,t)) ] \\&=&
  \sum_{m=0}^{m_0} [ p_{n,m}(1,t)m \log s -  p_{n,m} (1,t)\log(p_{n,m}(1,t)) ] \\&+& \sum_{m=m_0 + 1}^{n}
  [ p_{n,m}(1,t)m \log s - p_{n,m}(1,t) \log(p_{n,m}(1,t))) ]  \\&\equiv&
  \Sigma_1 + \Sigma_2.
\end{eqnarray*}
$\Sigma_1$ is bounded since it has a finite number $m_0$ of terms:
\begin{eqnarray*}
  \Sigma_1 &=&   \sum_{m=0}^{m_0} [ p_{n,m}m \log s -  p_{n,m} \log(p_{n,m}) ]  <
 \log(s) m_{0}^{2}-{m_{0}\over e}.
\end{eqnarray*}
where we used that $-p \log(p) < \frac{1}{e}$ for $0<p<1$.

To show that $\Sigma_2$ is bounded we use that by Lemma \ref{lem04}, $m>m_0 \Rightarrow p_{n,m}(1,t) \leq e^{-m}$, and that $-p \log(p) < -q \log (q)$ if $0 < p < q < \frac{1}{e}$.
With these relations we have
\begin{eqnarray*}
 \Sigma_2 &=& \sum_{m=m_0 + 1}^{n}
  [ p_{n,m} (1,t) m \log s - p_{n,m}(1,t) \log(p_{n,m}(1,t))) ] \\ &\leq& \sum_{m=m_0+1}^{\infty} e^{-m} (m \log s- \log(e^{-m})) = (1
  + \log s) \sum_{m=m_0 + 1}^{\infty} e^{-m} m\\ &\leq& (1 + \log s),
\end{eqnarray*}
where in the last step we used that $\sum_{m=1}^{\infty} e^{-m} m=\frac{e}{(e-1)^2} < 1$. 

We have thus established that the entropy is bounded by an $n$ independent constant, 
\begin{eqnarray}
  \label{eq:upperbound}
  S_n(s,t) < (m_0^2 + 1)\log(s) + \frac{m_0}{e} + 1~~.
\end{eqnarray}
\qed
 
\section{Conclusions}
The search for highly entangled ground states is one of the most important pursuits in the field of many-body entanglement. Precisely because these are not typical, yet are highly interesting and potentially useful. 

Here we took another step in this direction. We have introduced a novel highly entangled spin chain, the deformed Fredkin spin chain, and studied it's quantum phase diagram. The model's ground state has a simple interpretation as a superposition of weighted Dyck paths. In the colored case, it features an extensively entangled critical phase and an area law phase, while entanglement entropy of half the chain scales as a square root at the transition. 

Besides the high entanglement content of the model, its appeal is the relative simplicity of the interactions and the tractability of the ground state. The calculation of many more interesting quantities associated with the model is an ongoing project. These include behavior of correlations functions as function of $t$ and study of the gap, which is expected to be exponentially small in system size (a recent preprint, \cite{levine2016}, shows exponential gap scaling for the phase in \cite{zhang2016quantum}).

{\bf Acknowledgements:} We would like to thank A. Ahmadain, A. Feiguin and R. Movassagh for useful discussions. The work of IK and ZZ was supported by the NSF grant DMR-1508245. HK was supported in part by JSPS KAKENHI Grant No. JP15K17719 and No. JP16H00985. IK would like to thank the Simons Center for Geometry and Physics for hospitality where some of the work has been carried out.

\bibliographystyle{naturemag.bst}


\begin{thebibliography}{10}
\expandafter\ifx\csname url\endcsname\relax
  \def\url#1{\texttt{#1}}\fi
\expandafter\ifx\csname urlprefix\endcsname\relax\def\urlprefix{URL }\fi
\providecommand{\bibinfo}[2]{#2}
\providecommand{\eprint}[2][]{\url{#2}}

\bibitem{zhang2016quantum}
\bibinfo{author}{Zhang, Z.}, \bibinfo{author}{Ahmadain, A.} \&
  \bibinfo{author}{Klich, I.}
\newblock \bibinfo{title}{Quantum phase transition from bounded to extensive
  entanglement entropy in a frustration-free spin chain}.
\newblock \emph{\bibinfo{journal}{arXiv preprint arXiv:1606.07795}}
  (\bibinfo{year}{2016}).

\bibitem{heisenberg1928theorie}
\bibinfo{author}{Heisenberg, W.}
\newblock \bibinfo{title}{Zur theorie des ferromagnetismus}.
\newblock \emph{\bibinfo{journal}{Zeitschrift f{\"u}r Physik}}
  \textbf{\bibinfo{volume}{49}}, \bibinfo{pages}{619--636}
  (\bibinfo{year}{1928}).

\bibitem{bethe1931theorie}
\bibinfo{author}{Bethe, H.}
\newblock \bibinfo{title}{Zur theorie der metalle}.
\newblock \emph{\bibinfo{journal}{Zeitschrift f{\"u}r Physik}}
  \textbf{\bibinfo{volume}{71}}, \bibinfo{pages}{205--226}
  (\bibinfo{year}{1931}).

\bibitem{korepin1997quantum}
\bibinfo{author}{Korepin, V.}, \bibinfo{author}{Bogoliubov, N.~M.} \&
  \bibinfo{author}{Izergin, A.~G.}
\newblock \emph{\bibinfo{title}{Quantum inverse scattering method and
  correlation functions}}, vol.~\bibinfo{volume}{3}
  (\bibinfo{publisher}{Cambridge university press}, \bibinfo{year}{1997}).

\bibitem{beisert2012review}
\bibinfo{author}{Beisert, N.} \emph{et~al.}
\newblock \bibinfo{title}{Review of AdS/CFT integrability: an overview}.
\newblock \emph{\bibinfo{journal}{Letters in Mathematical Physics}}
  \textbf{\bibinfo{volume}{99}}, \bibinfo{pages}{3--32} (\bibinfo{year}{2012}).

\bibitem{affleck1987rigorous}
\bibinfo{author}{Affleck, I.}, \bibinfo{author}{Kennedy, T.},
  \bibinfo{author}{Lieb, E.~H.} \& \bibinfo{author}{Tasaki, H.}
\newblock \bibinfo{title}{Rigorous results on valence-bond ground states in
  antiferromagnets}.
\newblock \emph{\bibinfo{journal}{Physical review letters}}
  \textbf{\bibinfo{volume}{59}}, \bibinfo{pages}{799} (\bibinfo{year}{1987}).

\bibitem{affleck1988valence}
\bibinfo{author}{Affleck, I.}, \bibinfo{author}{Kennedy, T.},
  \bibinfo{author}{Lieb, E.~H.} \& \bibinfo{author}{Tasaki, H.}
\newblock \bibinfo{title}{Valence bond ground states in isotropic quantum
  antiferromagnets}.
\newblock \emph{\bibinfo{journal}{Comm. Math. Phys.}}
  \textbf{\bibinfo{volume}{115}}, \bibinfo{pages}{477--528}
  (\bibinfo{year}{1988}).

\bibitem{hastings2007area}
\bibinfo{author}{Hastings, M.~B.}
\newblock \bibinfo{title}{An area law for one-dimensional quantum systems}.
\newblock \emph{\bibinfo{journal}{Journal of Statistical Mechanics: Theory and
  Experiment}} \bibinfo{pages}{P08024} (\bibinfo{year}{2007}).

\bibitem{eisert2010colloquium}
\bibinfo{author}{Eisert, J.}, \bibinfo{author}{Cramer, M.} \&
  \bibinfo{author}{Plenio, M.~B.}
\newblock \bibinfo{title}{Colloquium: Area laws for the entanglement entropy}.
\newblock \emph{\bibinfo{journal}{Reviews of Modern Physics}}
  \textbf{\bibinfo{volume}{82}}, \bibinfo{pages}{277} (\bibinfo{year}{2010}).

\bibitem{holzhey1994high}
\bibinfo{author}{Holzhey, C.}, \bibinfo{author}{Larsen, F.} \&
  \bibinfo{author}{Wilczek, F.}
\newblock \bibinfo{title}{Geometric and renormalized entropy in conformal field
  theory}.
\newblock \emph{\bibinfo{journal}{Nucl. Phys. B424}}
  \textbf{\bibinfo{volume}{443}}, \bibinfo{pages}{467} (\bibinfo{year}{1994}).

\bibitem{calabrese2009entanglement}
\bibinfo{author}{Calabrese, P.} \& \bibinfo{author}{Cardy, J.}
\newblock \bibinfo{title}{Entanglement entropy and conformal field theory}.
\newblock \emph{\bibinfo{journal}{Journal of Physics A: Mathematical and
  Theoretical}} \textbf{\bibinfo{volume}{42}}, \bibinfo{pages}{504005}
  (\bibinfo{year}{2009}).

\bibitem{gioev2006entanglement}
\bibinfo{author}{Gioev, D.} \& \bibinfo{author}{Klich, I.}
\newblock \bibinfo{title}{{Entanglement entropy of fermions in any dimension
  and the Widom conjecture}}.
\newblock \emph{\bibinfo{journal}{Phys. Rev. Lett.}}
  \textbf{\bibinfo{volume}{96}}, \bibinfo{pages}{100503}
  (\bibinfo{year}{2006}).

\bibitem{wolf2006violation}
\bibinfo{author}{Wolf, M.}
\newblock \bibinfo{title}{{Violation of the entropic area law for fermions}}.
\newblock \emph{\bibinfo{journal}{Phys. Rev. Lett.}}
  \textbf{\bibinfo{volume}{96}}, \bibinfo{pages}{10404} (\bibinfo{year}{2006}).

\bibitem{gori2015explicit}
\bibinfo{author}{Gori, G.}, \bibinfo{author}{Paganelli, S.},
  \bibinfo{author}{Sharma, A.}, \bibinfo{author}{Sodano, P.} \&
  \bibinfo{author}{Trombettoni, A.}
\newblock \bibinfo{title}{Explicit hamiltonians inducing volume law for
  entanglement entropy in fermionic lattices}.
\newblock \emph{\bibinfo{journal}{Physical Review B}}
  \textbf{\bibinfo{volume}{91}}, \bibinfo{pages}{245138}
  (\bibinfo{year}{2015}).

\bibitem{irani2010ground}
\bibinfo{author}{Irani, S.}
\newblock \bibinfo{title}{Ground state entanglement in one-dimensional
  translationally invariant quantum systems}.
\newblock \emph{\bibinfo{journal}{Journal of Mathematical Physics}}
  \textbf{\bibinfo{volume}{51}}, \bibinfo{pages}{022101}
  (\bibinfo{year}{2010}).

\bibitem{gottesman2010entanglement}
\bibinfo{author}{Gottesman, D.} \& \bibinfo{author}{Hastings, M.}
\newblock \bibinfo{title}{Entanglement versus gap for one-dimensional spin
  systems}.
\newblock \emph{\bibinfo{journal}{New journal of physics}}
  \textbf{\bibinfo{volume}{12}}, \bibinfo{pages}{025002}
  (\bibinfo{year}{2010}).

\bibitem{vitagliano2010volume}
\bibinfo{author}{Vitagliano, G.}, \bibinfo{author}{Riera, A.} \&
  \bibinfo{author}{Latorre, J.}
\newblock \bibinfo{title}{Volume-law scaling for the entanglement entropy in
  spin-1/2 chains}.
\newblock \emph{\bibinfo{journal}{New Journal of Physics}}
  \textbf{\bibinfo{volume}{12}}, \bibinfo{pages}{113049}
  (\bibinfo{year}{2010}).

\bibitem{ramirez2014conformal}
\bibinfo{author}{Ram{\'\i}rez, G.}, \bibinfo{author}{Rodr{\'\i}guez-Laguna, J.}
  \& \bibinfo{author}{Sierra, G.}
\newblock \bibinfo{title}{From conformal to volume law for the entanglement
  entropy in exponentially deformed critical spin 1/2 chains}.
\newblock \emph{\bibinfo{journal}{Journal of Statistical Mechanics: Theory and
  Experiment}} \bibinfo{pages}{P10004} (\bibinfo{year}{2014}).

\bibitem{chen2014experimental}
\bibinfo{author}{Chen, M.}, \bibinfo{author}{Menicucci, N.~C.} \&
  \bibinfo{author}{Pfister, O.}
\newblock \bibinfo{title}{Experimental realization of multipartite entanglement
  of 60 modes of a quantum optical frequency comb}.
\newblock \emph{\bibinfo{journal}{Physical review letters}}
  \textbf{\bibinfo{volume}{112}}, \bibinfo{pages}{120505}
  (\bibinfo{year}{2014}).

\bibitem{Movassagh07112016}
\bibinfo{author}{Movassagh, R.} \& \bibinfo{author}{Shor, P.~W.}
\newblock \bibinfo{title}{Supercritical entanglement in local systems:
  Counterexample to the area law for quantum matter}.
\newblock \emph{\bibinfo{journal}{Proceedings of the National Academy of
  Sciences}}  (\bibinfo{year}{2016}).

\bibitem{salberger2016fredkin}
\bibinfo{author}{Salberger, O.} \& \bibinfo{author}{Korepin, V.}
\newblock \bibinfo{title}{Fredkin spin chain}.
\newblock \emph{\bibinfo{journal}{arXiv preprint arXiv:1605.03842}}
  (\bibinfo{year}{2016}).

\bibitem{dell2016violation}
\bibinfo{author}{Dell'Anna, L.}, \bibinfo{author}{Salberger, O.},
  \bibinfo{author}{Barbiero, L.}, \bibinfo{author}{Trombettoni, A.} \&
  \bibinfo{author}{Korepin, V.}
\newblock \bibinfo{title}{Violation of cluster decomposition and absence of
  light cones in local integer and half-integer spin chains}.
\newblock \emph{\bibinfo{journal}{Phys. Rev. B}} \textbf{\bibinfo{volume}{94}},
  \bibinfo{pages}{155140} (\bibinfo{year}{2016}).

\bibitem{bravyi2012criticality}
\bibinfo{author}{Bravyi, S.}, \bibinfo{author}{Caha, L.},
  \bibinfo{author}{Movassagh, R.}, \bibinfo{author}{Nagaj, D.} \&
  \bibinfo{author}{Shor, P.~W.}
\newblock \bibinfo{title}{Criticality without frustration for quantum spin-1
  chains}.
\newblock \emph{\bibinfo{journal}{Physical review letters}}
  \textbf{\bibinfo{volume}{109}}, \bibinfo{pages}{207202}
  (\bibinfo{year}{2012}).

\bibitem{andrews2004integer}
\bibinfo{author}{Andrews, G.~E.} \& \bibinfo{author}{Eriksson, K.}
\newblock \emph{\bibinfo{title}{Integer partitions}}
  (\bibinfo{publisher}{Cambridge University Press}, \bibinfo{year}{2004}).

\bibitem{levine2016}
\bibinfo{author}{Levine, L.} \& \bibinfo{author}{Movassagh, R.}
\newblock \bibinfo{title}{The gap of the area-weighted Motzkin spin chain is
  exponentially small}.
\newblock \emph{\bibinfo{journal}{arXiv preprint: 1611.03147}}
  (\bibinfo{year}{2016}).

\end{thebibliography}

\end{document}